\documentclass{article}
\usepackage[T1]{fontenc}
\usepackage{spconf,amsmath,graphicx,booktabs}
\usepackage{mathptmx} 
\usepackage[scaled=0.92]{helvet} 
\usepackage[hidelinks]{hyperref}
\usepackage{subcaption}
\usepackage{tikz}
\usetikzlibrary{positioning,arrows.meta,fit,calc}
\definecolor{oiblue}{HTML}{0072B2}   
\definecolor{oiorange}{HTML}{E69F00} 
\definecolor{oigreen}{HTML}{009E73}  
\usepackage{dblfloatfix} 

\usepackage{siunitx}
\usepackage{paralist}
\usepackage{microtype}

\graphicspath{{figures/}}

\title{Don't CLAP: Are Music-Text Models Bag-of-Words?}

\name{Yuan-Chiao Cheng \quad Alexander Lerch}
\address{Music Informatics Group, Georgia Institute of Technology}

\begin{document}

\maketitle

\begin{abstract}
Text-to-music systems are assessed on audio quality and on how faithfully the music follows its prompt, and the CLAP score, the cosine similarity between a music-text model's audio and text embeddings, is the standard objective metric of faithfulness.
We ask how accurately that score reflects the text: when an attribute is linked to an instrument (e.g., \emph{distorted} guitar), does the text embedding capture that binding?
To find out, we introduce an attribute swap perturbation: the caption of a real recording is edited by exchanging exactly one property, timbre, lead versus accompaniment, or order of first appearance, between two instruments. We then test four contrastive music-text models and one large audio-language model on whether the audio scores higher against the original caption than against the perturbed one.
No contrastive model distinguishes the two captions reliably. The audio-language model does better, but further experiments show that its advantage rests largely on audio-agnostic language priors.
Our results thus provide compelling evidence that the CLAP score and related metrics do not capture fine-grained musical meaning or attribute bindings; their representation is closer to a bag-of-words that leaves them insensitive to meaning-changing perturbations of the caption.
\end{abstract}


\section{Introduction}

    The two most common evaluation targets for text-to-music systems are audio quality and prompt faithfulness \cite{lerch_survey_2025}. Audio quality assessment is often attempted through proxy metrics such as FAD and KAD \cite{fad,kad}, while prompt faithfulness, i.e., how closely the music characteristics reflect the text prompt, has been most frequently measured subjectively with listening studies \cite{musicarena} or objectively with the CLAP score, the cosine similarity between the audio and text embeddings of a contrastive music-text model \cite{laionclap,msclap,musicgen,musicldm}. Although large audio-language models (LALMs) are capable of writing fluent music descriptions \cite{audioflamingo3,qwen2audio}, they seem to answer music questions largely from text priors \cite{rulistening}, which is one reason why the CLAP score remains the predominant metric \cite{musicgen,audioldm2,musicldm}.

    CLAP models inherit their two-encoder contrastive architecture from CLIP \cite{clip}, whose weaknesses are well documented in computer vision. Yuksekgonul et al.\ show that CLIP behaves like a bag-of-words: it matches an image to the concepts a caption names but ignores their attributes, relations, and order \cite{aro,winoground}. More recent results indicate that the text and image embeddings each still encode which attribute belongs to which object, and that this information is only lost in the cross-modal comparison \cite{labclip}.

    Our study also draws inspiration from prior work in the audio domain. Ghosh et al.\ build paired audio-caption sets differing only in the order or attributes of environmental sound events and find contrastive audio-language models to perform barely above chance level at picking the matching caption \cite{compa}. Zang et al.\ show that LALMs answer music question-answering benchmarks from text priors, scoring well even without the audio \cite{rulistening}. It is unclear, however, whether music-text models share the bag-of-words shortcoming; if they do, a caption such as ``a distorted guitar over a clean piano'' would be indistinguishable from ``a clean guitar over a distorted piano,'' which obviously could have major implications for the musical meaning of the text embedding.

    Simply shuffling caption words does not test this, as many shuffles are musically equivalent (e.g., ``lo-fi C minor beat'' vs.\ ``C minor lo-fi beat''). We therefore perturb each caption by swapping exactly one property between the two instruments of a recording while keeping all other words, for three attribute dimensions: timbre, lead voice versus accompaniment, and order of first appearance. With the audio unchanged, a sample counts as correct when the audio is more similar to the original caption $c^{+}$ than to the perturbed caption $c^{-}$. A text embedding perfectly encoding the word order would thus score 100\% accuracy, while a text embedding ignoring the word order would score at 50\% chance level.

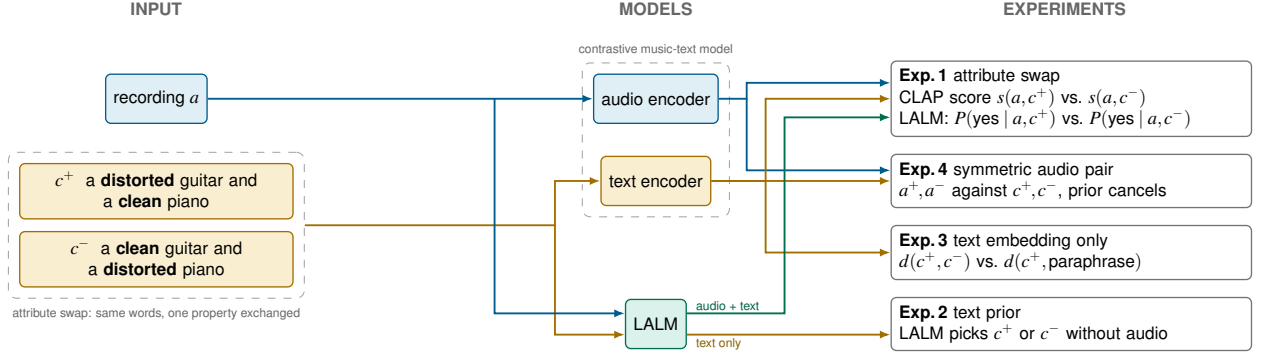
\begin{figure*}[t]
\centering
\begin{tikzpicture}[font=\sffamily\scriptsize, >={Latex[length=1.4mm, width=1.1mm]}, x=1cm, y=1cm,
  every node/.style={align=center},
  hdr/.style={font=\sffamily\scriptsize\bfseries, text=black!60, anchor=south},
  box/.style={draw=black!55, line width=0.5pt, rounded corners=2pt, inner sep=3pt, minimum height=6.5mm, fill=white},
  aud/.style={box, fill=oiblue!12, draw=oiblue!80!black},
  txt/.style={box, fill=oiorange!18, draw=oiorange!70!black},
  lalm/.style={box, fill=oigreen!14, draw=oigreen!70!black},
  exp/.style={box, fill=white, draw=black!55, anchor=north west, text width=4.55cm, align=left, inner ysep=3pt},
  grp/.style={draw=black!35, dashed, rounded corners=3pt, inner sep=4pt},
  arrA/.style={->, draw=oiblue!80!black, line width=0.7pt},
  arrT/.style={->, draw=oiorange!70!black, line width=0.7pt},
  arrL/.style={->, draw=oigreen!70!black, line width=0.7pt}]
\coordinate (colI) at (1.75,0);
\coordinate (colM) at (8.35,0);
\node[exp] (e1) at (11.45,1.78) {\textbf{Exp.\,1} attribute swap\\ CLAP score $s(a,c^{+})$ vs.\ $s(a,c^{-})$\\ LALM: $P(\text{yes}\mid a,c^{+})$ vs.\ $P(\text{yes}\mid a,c^{-})$};
\node[exp] (e4) at ($(e1.south west)+(0,-0.22)$) {\textbf{Exp.\,4} symmetric audio pair\\ $a^{+}\!,a^{-}$ against $c^{+}\!,c^{-}$, prior cancels};
\node[exp] (e3) at ($(e4.south west)+(0,-0.22)$) {\textbf{Exp.\,3} text embedding only\\ $d(c^{+},c^{-})$ vs.\ $d(c^{+},\text{paraphrase})$};
\node[exp] (e2) at ($(e3.south west)+(0,-0.22)$) {\textbf{Exp.\,2} text prior\\ LALM picks $c^{+}$ or $c^{-}$ without audio};
\node[aud] (aenc) at (colM |- e1.center) {audio encoder};
\node[txt] (tenc) at (colM |- e4.center) {text encoder};
\node[grp, fit=(aenc)(tenc), label={[font=\sffamily\tiny, text=black!60]above:contrastive music-text model}] (clap) {};
\node[lalm] (lalm) at (colM |- e2.center) {LALM};
\node[aud] (rec) at (colI |- e1.center) {recording $a$};
\node[txt, text width=3.4cm] (pos) at ($(1.75,0)+(0,0.05)$) {$c^{+}$\; a \textbf{distorted} guitar and\\ a \textbf{clean} piano};
\node[txt, text width=3.4cm] (neg) at ($(1.75,0)+(0,-0.85)$) {$c^{-}$\; a \textbf{clean} guitar and\\ a \textbf{distorted} piano};
\node[grp, fit=(pos)(neg), label={[font=\sffamily\tiny, text=black!60]below:attribute swap: same words, one property exchanged}] (cap) {};
\node[hdr] at (1.75,2.25) {INPUT};
\node[hdr] at (8.35,2.25) {MODELS};
\node[hdr] at (13.75,2.25) {EXPERIMENTS};
\draw[arrA] (rec.east) -- (aenc.west);
\coordinate (ba) at ($(aenc.west)+(-1.3,0)$);
\coordinate (bt) at ($(tenc.west)+(-0.6,0)$);
\draw[arrT, preaction={draw=white, line width=2.4pt}] (cap.east) -- (cap.east -| bt) |- (tenc.west);  
\draw[arrT] (cap.east -| bt) |- ([yshift=-4pt]lalm.west);
\draw[arrA] (ba |- rec.east) |- ([yshift=4pt]lalm.west);
\coordinate (oa) at (9.55,0);
\coordinate (ot) at (9.8,0);
\coordinate (ol) at (10.05,0);
\draw[arrA] (aenc.east) -- (aenc.east -| oa) |- ([yshift=6pt]e1.west);
\draw[arrA] (aenc.east -| oa) |- ([yshift=4pt]e4.west);
\draw[arrT] (tenc.east) -- (tenc.east -| ot) |- (e1.west);
\draw[arrT] (tenc.east) -- (tenc.east -| e4.west);
\draw[arrT] (tenc.east -| ot) |- (e3.west);
\draw[arrL] ([yshift=4pt]lalm.east) -- ([yshift=4pt]lalm.east -| ol) |- ([yshift=-7pt]e1.west);
\node[font=\sffamily\tiny, text=oigreen!60!black, anchor=south west, inner sep=1pt] at ($([yshift=4pt]lalm.east)+(0.08,0)$) {audio + text};
\draw[arrT] ([yshift=-4pt]lalm.east) -- ([yshift=-4pt]lalm.east -| e2.west);
\node[font=\sffamily\tiny, text=oiorange!60!black, anchor=north west, inner sep=1pt] at ($([yshift=-4pt]lalm.east)+(0.08,0)$) {text only};
\end{tikzpicture}
\caption{Overview: a recording $a$ and its caption pair ($c^{+}$ original, $c^{-}$ swapped) are scored by contrastive music-text models and a LALM in the four experiments.}
\label{fig:flow}
\end{figure*}

    The contributions of the paper are threefold:
    First, we present the \textbf{Music Attribute-Swap Benchmark (MASB)}: 400 annotated recordings with attribute-swap caption pairs for the three properties, plus MASB-Order, a symmetric audio set that cancels the text prior.
    Second, we demonstrate that current contrastive \textbf{music-text models cannot distinguish original from perturbed captions} for a given audio input (Exp.~1) and that the embeddings of the original and perturbed captions can barely be separated (Exp.~3), supporting our bag-of-words hypothesis. 
    Third, we provide evidence that, although our LALM outperforms the contrastive models, this \textbf{LALM advantage is largely due to the language prior}: on lead versus accompaniment and onset order its performance tracks the choices it makes from the captions alone, without audio (Exp.~2), and it performs only at near-chance level on a carefully constructed prior-free audio subset (Exp.~4).

\section{Methodology}
\label{sec:method}
    We ask whether a music-text model's text embedding preserves the attribute relations in a caption.
    Figure~\ref{fig:flow} outlines the four experiments that answer this.

\subsection{Dataset}
\label{sec:mbb}

    A recording is an unmodified \SI{10}{\second} excerpt from a Creative-Commons track in the Song Describer Dataset \cite{songdescriber} or MTG-Jamendo \cite{mtgjamendo}, cut by a single trained annotator to ensure consistency and reduce annotation ambiguity. The clip is annotated with the names of the two instruments and with one property: a free-form timbre adjective per instrument, the role of each instrument (lead or accompaniment), or their order of first appearance in the audio.
    Each recording is annotated for exactly one property, which gives 140 timbre, 123 lead-versus-accompaniment, and 137 onset-order recordings, resulting in an overall set of 400 audio clips.
    Each clip is annotated in two caption formats, one following a strict template and the other a natural rewrite of the same content, for example ``the guitar sounds distorted while the piano is clean.'' This leads to 800 caption pairs.

    In addition, another subset is added for Exp.~4: MASB-Order. It contains 300 symmetric pairs from MoisesDB stems \cite{moisesdb}: mixes $a^{+}$ and $a^{-}$ of one track differ only in which of two instruments enters first (the later one silent for the first \SI{5}{\second}), and captions $c^{+}$ ``A enters before B'' and $c^{-}$ ``B enters before A'' are each correct for one of the two mixtures. We release the identifiers, annotations, captions, and code.\footnote{\url{https://github.com/barry-mir/music-clap-bow}}

\begin{figure*}[t!]
\begin{minipage}[t]{0.49\textwidth}
\vspace{0pt}
\centering
\includegraphics[width=\linewidth]{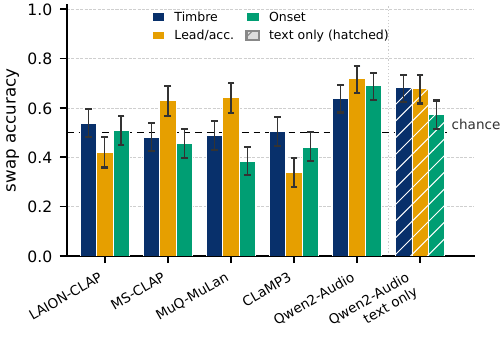}
\captionof{figure}{Exp.\,1: swap accuracy per model and property, with 95\% CIs and the chance line (hatched: LALM without audio).}
\label{fig:swapacc}
\end{minipage}\hfill
\begin{minipage}[t]{0.49\textwidth}
\vspace{0pt}
\centering
\includegraphics[width=\linewidth]{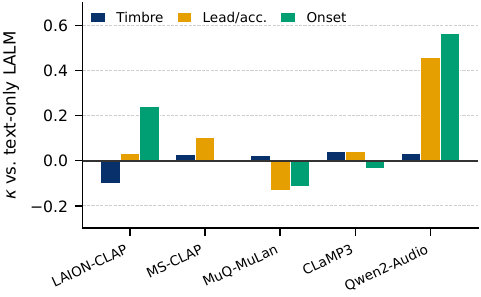}
\captionof{figure}{Exp.\,2: agreement (Cohen's $\kappa$) between each model's choices and the LALM's text-only choices.}
\label{fig:agree}
\end{minipage}
\end{figure*}

\subsection{Caption construction and perturbation}
\label{sec:props}

    Each recording is described by two instruments and a tag characterizing one property for each instrument. For the \textbf{timbre} property, the annotator describes each instrument with one free-form adjective, such as \texttt{distorted}, \texttt{clean}, \texttt{bright}, or \texttt{warm}; the two adjectives of a recording always differ. For the \textbf{role} property, the annotation simply flags which instrument plays the melody (lead), and which the accompaniment. For the \textbf{order} property, we log which instrument appears first.

    The captions are constructed based on templates. The timbre template is ``a \texttt{<timbre A>} \{instrument A\} and a \texttt{<timbre B>}  \{instrument B\}''. The instrument role template is ``the \{lead instrument\} plays the melody while the \{accompaniment instrument\} accompanies''. The order template is ``the \{first instrument\} enters before the \{second instrument\}''.
    Each recording is additionally rendered with a second, more natural template carrying the same content: ``the \{instrument A\} sounds \texttt{<timbre A>} while the \{instrument B\} is \texttt{<timbre B>}'', ``the \{lead\} leads with the tune as the \{accompaniment\} backs it'', and ``first the \{first\} comes in, then the \{second\}''; all results pool both templates.

    The attribute swap exchanges the two instruments' slots inside the template, so the original caption $c^{+}$ and the swapped caption $c^{-}$ contain the exact same words.


\subsection{Models and scoring}
\label{sec:task}

    To ensure that our results do not only hold true for one specific model or architecture but generalize across a variety of current models, we examine four contrastive music-text models that differ in text encoder and training data: the LAION-CLAP music checkpoint \cite{laionclap}, MS-CLAP 2023 \cite{msclap}, MuQ-MuLan \cite{muq}, and CLaMP 3 \cite{clamp3}.
    As an additional reference, we add a LALM, Qwen2-Audio-7B-Instruct \cite{qwen2audio}, which we run both with the audio (Exp.~1) and, as its own text-only baseline, without it (Exp.~2).
    
    For each audio clip $a$ and caption pair $(c^{+}, c^{-})$, a model is considered correct when the cosine similarity $s(\cdot,\cdot)$ between the audio embedding and the original caption embedding (the CLAP score) is larger than that between the audio embedding and the perturbed caption embedding, i.e., $s(a, c^{+}) > s(a, c^{-})$; the chance accuracy is 0.5.

    In case of the LALM, the model is asked, given the audio and one caption: ``Does this description match the audio? Answer Yes or No.'' The fraction of correct answers gives the LALM accuracy.
    In the text-only setting, the LALM sees both captions and picks one without seeing the audio, read from the next-token probabilities of the two option labels. The answer is averaged over the two caption orderings to cancel position bias; its accuracy is the floor that language priors alone reach without any audio information.
    
    Where a distance between embeddings is needed we use the cosine distance.
    We report 95\% confidence intervals (CIs) and exact binomial $p$-values, Holm-corrected over all tests.

\subsection{Separating text prior from audio evidence}
\label{sec:prior-method}

    A caption such as ``a distorted guitar and a clean piano'' is more typical than its swapped version, so a model may pick it from the text alone, with or without audio. We separate this prior from audio evidence in two ways.
    First, Cohen's $\kappa$ \cite{cohen1960} measures how much a model's choices with audio agree with its own text-only choices beyond what their accuracies alone would predict ($\kappa=0$: no agreement beyond chance, $\kappa=1$: identical choices).
    Second, following the blind-baseline logic of prior work \cite{sugarcrepe,priorwu}, we split the caption pairs by whether the text-only choice is correct. By Bayes' rule, the decision is the text prior plus the audio evidence \cite{good1950}. The split changes only the prior, so a model that reads the audio scores alike on both halves, and a model that follows the prior scores high on one and low on the other. We remove the prior's advantage with prior-balanced accuracy, the mean of the two halves.

\section{Experiments}
\label{sec:experiments}

    We run four experiments. Exp.~1 evaluates the audio-caption similarity by assessing the correctness of the caption pair with higher similarity. Exp.~2 investigates the role of the language prior in that decision by comparing the LALM's choices with and without audio. Exp.~3 examines the text embedding proximity by comparing the attribute swapped proximity with that of a paraphrased caption. Finally, Exp.~4 removes the text prior by construction on a symmetric audio set.

\subsection{Exp.\,1: CLAP score under attribute swap}
\label{sec:diagnosis}

\begin{figure*}[t!]
\begin{minipage}[t]{0.49\textwidth}
\vspace{0pt}
\centering
\includegraphics[width=\linewidth]{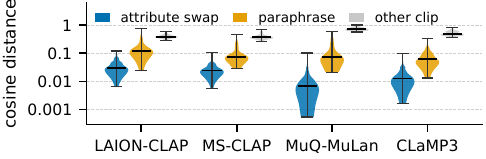}
\captionof{figure}{Exp.\,3: per-pair cosine distance from $c^{+}$ to the swapped caption, a paraphrase, and the median over all other recordings' captions; ticks mark medians, log $y$ axis.}
\label{fig:textenc}
\end{minipage}\hfill
\begin{minipage}[t]{0.49\textwidth}
\centering\small
\vspace{0pt}
\captionof{table}{Exp.\,4: accuracy on the symmetric audio set MASB-Order, with 95\% CIs; chance is 0.5.}
\label{tab:order}
\setlength{\tabcolsep}{6pt}
\begin{tabular}{lcc}
\toprule
Model & Accuracy & [95\% CI] \\
\midrule
LAION-CLAP & 0.508 & [0.468, 0.548] \\
MS-CLAP & 0.538 & [0.498, 0.578] \\
MuQ-MuLan & 0.497 & [0.457, 0.537] \\
CLaMP 3 & 0.527 & [0.487, 0.566] \\
Qwen2-Audio & 0.537 & [0.497, 0.576] \\
\bottomrule
\end{tabular}
\end{minipage}
\end{figure*}

    Fig.~\ref{fig:swapacc} plots the swap accuracy, the fraction of caption pairs with $s(a, c^{+}) > s(a, c^{-})$, per model and property; the four contrastive models show no consistent preference for the original caption.
    The mean accuracy across properties ranges from 0.431 to 0.519. Two above-chance results stand out after Holm correction at adjusted $p<0.05$, both on lead role: MuQ-MuLan (0.642) and MS-CLAP (0.630). Curiously, CLaMP 3 yields an accuracy of 0.337 for the same property and MuQ-MuLan 0.383 on onset order, both considerably below chance.
    No model stays above chance on all three properties, and even the two statistically significant above-chance results have far too low accuracy to be called reliable. Thus, no model can claim prompt faithfulness under caption perturbations.

    The LALM Qwen2-Audio outperforms the other models: its mean accuracy reaches 0.681, and all three results survive Holm correction, timbre (0.639), lead role (0.720), and onset order (0.690).


\subsection{Exp.\,2: Text prior versus audio evidence}
\label{sec:agree}

    Following Hsieh et al., who add a text-only baseline to such benchmarks because a model often simply picks the more plausible caption and thus looks better than it actually is \cite{sugarcrepe}, we first score the LALM on the captions alone. Without audio it already reaches 0.682 on timbre and 0.679 on lead role, above every contrastive model; on onset order it stays close to chance at 0.573 ($p=0.018$), since text carries little prior for which instrument starts first. Adding the audio raises the overall accuracy only from 0.644 to 0.681, and on timbre it even lowers it, from 0.682 to 0.639.
    Figure~\ref{fig:agree} shows that the four contrastive models never exceed $\kappa=0.24$ and mostly stay near zero, so their choices are unrelated to the language prior.
    Qwen2-Audio with audio reaches $\kappa=0.46$ with its own text-only choices on lead role, $0.57$ on onset order, and $0.03$ on timbre. Diving into the detailed results, we note that on the pairs where its text-only choice is wrong, it picks the correct caption on only 0.42 of lead role pairs, far below the 0.86 it reaches where that choice is right, and on 0.38 of onset-order pairs against 0.92; on timbre the two halves are alike, 0.62 against 0.65.
    Averaging the two accuracies, prior-balanced accuracy is 0.64 on lead role, 0.65 on onset order, and 0.63 on timbre. Thus, on lead role and onset order the audio mostly amplifies the choice the model would make from the captions alone, while on timbre a small margin remains that the prior does not explain.

\subsection{Exp.\,3: Single-modality text embedding proximity}
\label{sec:localization}

    Exp.\,3 asks whether the text embedding by itself separates $c^{+}$ from $c^{-}$ for the four contrastive models (Qwen2-Audio has no joint embedding space).
    We embed $c^{+}$ and compute the cosine distance to three references: the perturbed caption $c^{-}$ (identical words); a paraphrase built from a fixed synonym map (identical meaning); and, as an upper reference, the median distance to the original captions of all other recordings.
    An encoder that keeps instrument relations should move farther under the attribute swap, which changes the meaning, than under the paraphrase, which keeps it.
    
    Figure~\ref{fig:textenc}, however, shows the opposite for every encoder: the median distance to the swapped caption is 0.007 to 0.030, below the paraphrase at 0.062 to 0.122 and far below other recordings' captions at 0.37 to 0.71.
    We can therefore conclude that the failure point is already present in the text embedding itself before any audio is compared.
    Two properties of contrastive training may explain this, though neither is tested here.
    First, the contrastive objective never asks the model to reproduce a caption; separating the paired caption from in-batch negatives rarely requires parsing every detail, incentivizing to disregard word order at no cost.
    Second, much music-text training data pairs audio with tags or short phrases, not full sentences, which gives the text encoder little long term context to model.

\subsection{Exp.\,4: Symmetric audio onset-order test}
\label{sec:audio}

    Exp.~1 to 3 perturb the caption and not the audio, because re-rendering one instrument's timbre or role would be an invasive change that may alter other properties of the recording.
    MASB-Order (Section~\ref{sec:mbb}) constructs a mirror pair for each audio sample, which allows two comparisons per pair: $a^{+}$ is counted correct if it scores $c^{+}$ above $c^{-}$, and $a^{-}$ if it scores $c^{-}$ above $c^{+}$.
    Thus, we construct a prior-free audio-caption subset with a true chance level of 0.5.
    Table~\ref{tab:order} shows the results for these data: every model performs close to chance from 0.497 to 0.538. That means that with the prior removed, none reads from the audio which instrument starts first.
    We note that Qwen2-Audio's prior-balanced onset-order margin from Exp.~2 does not carry over to this experiment, but we should also clarify that the audio content of the two sets differ: MASB-Order comes from MoisesDB stems with a muting edit and template captions, while MASB comprises real recordings with human annotation, so a distribution shift might explain the gap.



\section{Conclusion}
    Our experimental results confirm that the caption in contrastive music-text models is indeed embedded like a bag-of-words.
    Four contrastive models rank the swapped caption at chance (Exp.~1), and the attribute swap barely moves their text embedding (Exp.~3).
    The LALM Qwen2-Audio scores higher, but its main competitive advantage stems from the text prior (Exp.~2, 4).
    While the CLAP score certainly will remain an important assessment tool, our results indicate that caution is prudent before assuming faithful text encoding of long form, musically detailed captions.

\ninept
\bibliography{references}

\end{document}